\documentclass{aa}
\usepackage{psfig}
\usepackage{latexsym}
\usepackage{longtable}
\usepackage{array}
\begin{document}
    \title{Asiago eclipsing binaries program. II. V505 Per\thanks{based partly 
          on data obtained with Asiago 1.82 m telescope}$^{,}$\thanks{ 
          Table~3 available in electronic form only}}

\author	{L. Tomasella \inst{1}
	\and
	U. Munari\inst{1}
	\and
	A. Siviero\inst{1}
	\and
	S. Cassisi\inst{2}
	\and
	S. Dallaporta\inst{3}
	\and
	T. Zwitter\inst{4}
	\and
	R. Sordo\inst{1}
          }

\institute{
	INAF Osservatorio Astronomico di Padova, Sede di Asiago, 36012
	Asiago (VI), Italy
	\and
	INAF Osservatorio Astronomico di Collurania, Via M. Maggini, 64100 Teramo, Italy
        \and
        via Filzi 9, 38034 Cembra (TN), Italy
	\and
	University of Ljubljana, Department of Physics, Jadranska 19, 1000 Ljubljana, Slovenia
        }

   \date{Received date ................; accepted date ...............}

\abstract{The orbit and fundamental physical parameters of the double-lined
eclipsing binary V505~Per are derived by means of Echelle high resolution,
high S/N spectroscopy and $B$, $V$ photometry. In addition, effective
temperatures, gravities, rotational velocities and metallicities of both
components are obtained also from atmospheric $\chi^2$ analysis, showing
excellent match with the results of orbital solution. An
$E_{B-V}\leq$~0.01 mag upper limit to the reddening is derived from
intensity analysis of interstellar NaI (5890.0 \& 5895.9 \AA) and KI (7699.0
\AA) lines. The distance to the system computed from orbital parameters
(60.6$\pm$1 pc) is identical to the newly re-reduced Hipparcos parallax
(61.5$\pm$1.9 pc.). The masses of the two components
(M$_1$=1.2693$\pm$0.0011 and M$_2$=1.2514$\pm$0.0012 M$_\odot$) place them
in the transition region between convective and radiative stellar cores of
the HR diagram, with the more massive of the two showing already the effect
of evolution within the Main Sequence band (T$_1$=6512$\pm$21~K,
T$_2$=6462$\pm$12~K, R$_1$=1.287$\pm$0.014, R$_2$=1.266$\pm$0.013
R$_\odot$). This makes this system of particular relevance to theoretical
stellar models, as a test on the overshooting. We compare the firm
observational results for V505 Per component stars with the predictions of
various libraries of theoretical stellar models (BaSTI, Padova, Granada,
Yonsei-Yale, Victoria-Regina) as well as BaSTI models computed
specifically for the masses and chemical abundances of V505~Per. We found
that the overshooting at the masses of V505 Per component stars is already
pretty low, but not null, and described by efficiencies $\lambda_{\rm
OV}=0.093$ and 0.087 for the 1.27 and 1.25 M$_\odot$ components,
respectively. According to the computed BaSTI models, the age of the system
is $\sim$0.9~Gyr and the element diffusion during this time has reduced the
surface metallicity from the initial [M/H]=$-$0.03 to the current
[M/H]=$-$0.13, in excellent agreement with observed [M/H]=$-$0.12$\pm$0.03.

     \keywords{stars: fundamental parameters --
                binaries: spectroscopic --
                binaries: eclipsing -- star: individual: V505 Per}
            }

   \maketitle

\section{Introduction}

Double-lined eclipsing binaries (SB2 EBs) represent a primary tool to
provide fundamental stellar parameters, first of all masses and radii. These
parameters, when measured with high accuracy, represent a formidable
benchmark for the current generation of stellar evolutionary models. In
fact, they have to simultaneously fit the two stars of the binary, by
adopting exactly the same age and the same initial chemical composition 
for them.

On the other hand, since the reliability of stellar models is still
partially hampered by our poor knowledge of some physical processes at work
in real stars, such as ($a$) the efficiency of core convective overshoot
during the core H-burning phase in intermediate mass stars, (i.e. stars with
mass $M\ge 1.1-1.2$~M$_\odot$, the exact value depending on the chemical
composition), ($b$) how the efficiency of core convective overshoot
decreases with decreasing stellar mass, in the mass range where the
transition between fully convective to fully radiative stellar cores occurs,
or ($c$) the efficiency of convection in the super-adiabatic layers (cf.
Ribas, Jordi \& Gim{\'e}nez 2000), observational data from SB2 EBs provide
strong constraints for the different approaches used in stellar model
computations. In this context, the most suitable type of binary systems are
those with the mass of at least one of the two component is of the order of
$\sim1.2$~M$_\odot$ (cf. Pietrinferni et al. 2004, hereafter P04). 
V505~Per is such a system, according to earlier determinations of the 
masses (i.e. Marschall et al. 1997, hereafter M97; Munari et al. 2001, 
hereafter M01).

In order to properly use binary systems to constrain the accuracy of
current stellar evolutionary models, the properties of both components
should be known at the 1-2\% level and should be accompanied by accurate
determination of effective temperatures and metallicities (which are not a
direct product of orbital solution).

Andersen (1991, 1997, 2002) listed about fifty eclipsing binary systems for
which fundamental stellar parameters at the level of 1-2\% have been obtained
so far in the literature from the modeling of their orbits. They have been
compared to predictions of stellar evolutionary models by Pols et al.
(1997). Andersen (2002) argued in favor of more systems to be observed and
modeled at this level of accuracy, and Pols et al. (1997) stressed how
unknown metallicity for the majority of these systems with excellent orbits
spoiled the potential of comparison to theoretical predictions.

In the present series of papers, fundamental physical parameters for SB2 EBs
are derived by means of high precision photometric and spectroscopic data,
as well as accurate orbital solutions via the Wilson-Devinney code (Wilson
\& Devinney 1971, Wilson 1998, Milone et al. 1992).  Contrary to common
practice, we do not assume a temperature for primary star (deduced for
example from photometric colors or spectral types) and the reddening
affecting the binary, but we measure them both. Reddening comes from direct
measurements of the interstellar NaI (5890 \& 5896 \AA) and KI (7665 \& 7699
\AA) doublets, using the widely-used relation between equivalent width and
color excess $E_{B-V}$ calibrated by Munari \& Zwitter (1997). In addition,
we perform an atmospheric analysis of both stars in the binary system by
means of $\chi^2$ fit to the Kurucz synthetic spectral library computed by
Munari et al. (2005, hereafter M05) at the same $R$=20\,000 resolving power 
of Echelle spectra used in this series of papers. The $\chi^2$ fit provides
temperature, gravity, metallicity and projected rotational velocity for both
components. The M05 synthetic spectral library is the same as adopted by the 
RAVE project for analysis of its digital spectroscopic
survey over the whole southern sky (Steinmetz et al. 2006).

\section{V505~Per}

  \begin{table}  
    \caption{Radial velocities (km sec$^{-1}$) of V505~Per. The columns give the spectrum
    number (from the Asiago Echelle log book), the heliocentric JD
    ($-$2450000), the orbital phase, the radial velocities of the two
    components and the corresponding errors, and the $<$S/N$>$ of the
    spectrum averaged over the wavelength range considered in the analysis.}
    \centerline{\psfig{file=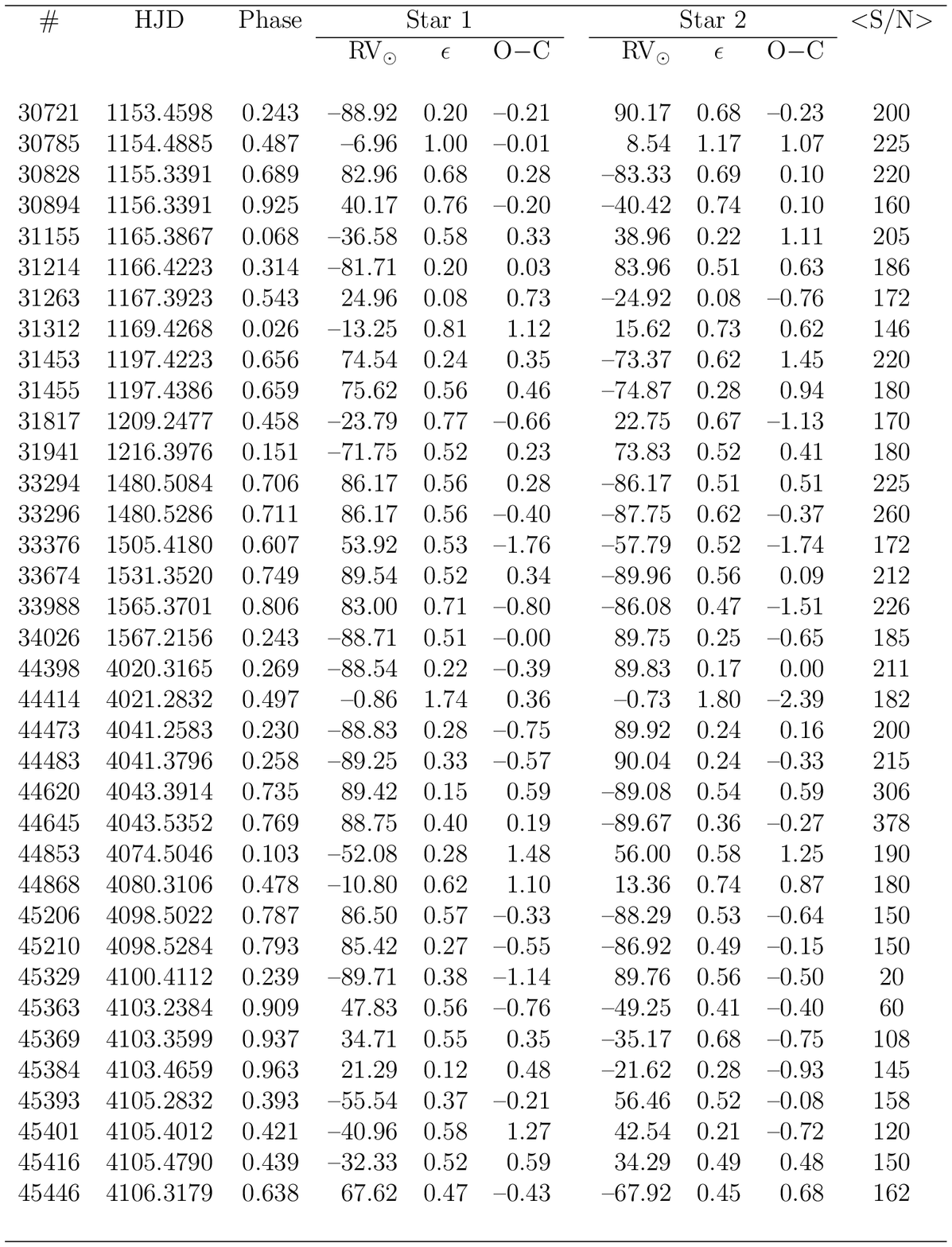,width=8.7cm}}
  \end{table}

In this paper we deal with V505~Per (HD 14384, HIP 10961), a nearby
eclipsing binary with two F5V components, discovered by Kaiser (1989) and
first studied by Kaiser et al. (1990), who suggested an orbital period
$\sim$2.1 days, and by Marschall et al. (1990) who argued in favor of an
orbital period twice longer.

M97 obtained the first spectroscopic and
photometric full orbital solution, with formal errors on radii and masses of
2.5\% and 1\%, respectively. No atmospheric analysis was performed, the
temperature of the primary was adopted from the spectral type and the
reddening was assumed to be zero given the proximity of the star.

A second, independent orbital study was performed by M01 while estimating
the performances of the ESA's GAIA satellite. They adopted Hipparcos
photometry as a suitable approximation for GAIA photometry, and supplemented
by ground-based spectroscopy in the same wavelength rage and similar
resolving power as planned at that time for the satellite. They derived the
temperature of the primary from Tycho
$(B-V)_{\rm T}$ color, assuming again a zero reddening. No atmospheric
analysis was performed neither.

These two orbital solutions are in reasonable agreement, although radii and
masses differ by more than the combined respective formal errors.  An
independent, full scale new investigation of V505~Per is therefore in order,
and to achieve it we proceeded with acquisition of a new, independent and
complete set of both photometric and spectroscopic data. In addition to the
orbital solution carried out simultaneously on the light and radial velocity
curves, we also provide atmospheric analysis and a direct
reddening determination via interstellar lines.

\section{The data}

\subsection{Photometry}

Similarly to Paper~I in this series (Siviero et al. 2004),  the photometric
observations of V505~Per were obtained with a 28 cm Schmidt-Cassegrain
telescope equipped with an Optec SSP5 photoelectric photometer.  In all, the
measurements for V505 Per consist of 311 points in $B_{\rm J}$ and 316 in
$V_{\rm J}$ (standard Johnson bands), secured between 2000 and 2001 and well
covering the whole lightcurve. The comparison star was HD 14444 (F0 spectral
type) and the check star was HD 14062 (K0 spectral type).  Their adopted
values were $V_{\rm J}$=8.760, $(B-V)_{\rm J}$=+0.432 and $V_{\rm J}$=7.658,
$(B-V)_{\rm J}$=+1.057, respectively. They were derived from Tycho-2 $B_{\rm
T}$, $V_{\rm T}$ photometry ported to Johnson's system by means of Bessell
(2000) transformations. Both comparison and check stars lay close to
V505~Per on the sky (16.42 and 28.16 arcmin, respectively) and share similar
colors, so that differential atmospheric absorption is not a source of
uncertainty in our photometry. The comparison star was measured against the
check star at least once every observing run, and found stable at better
than 0.01 mag level.

The whole set of photometric data is reported in Table~3 (available
electronic only).  The $B_{\rm J}$, $V_{\rm J}$ lightcurves and $(B-V)_{\rm
J}$ color curve are plotted in Fig.~2. They do not show hints of intrinsic
variability of any of the two stars. The dispersion of $B_{\rm J}$ and
$V_{\rm J}$ points around the orbital solution plotted in Fig.~2 is
$\sigma_B$=0.008 and $\sigma_V$=0.007, respectively, arguing in favor of a 
high internal consistency of our photometric data.

    \begin{figure}
    \centerline{\psfig{file=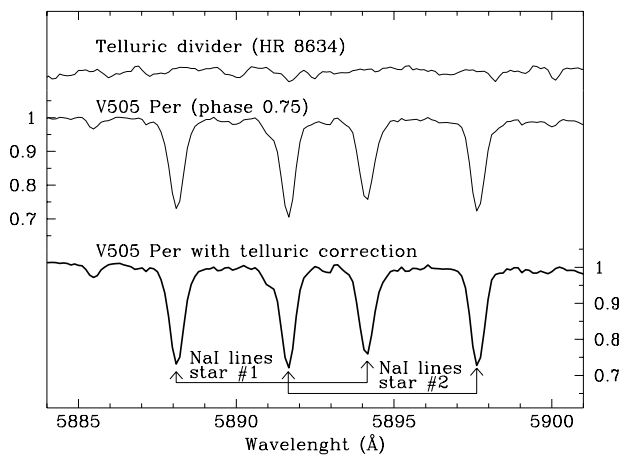,width=8.5cm,angle=270}}
    \caption{A zoomed view of the NaI doublet (5890 \& 5896 \AA) region for
        V505~Per at orbital phase 0.75 from an Asiago Echelle spectrum.  The
        observed V505 Per spectrum is at center. On top there is the
        spectrum of a fast rotating, unreddened B-type star observed
        immediately after the binary at similar airmass to serve as a mapper
        of the telluric absorptions. At the bottom there is the V505 Per
        after being divided by the B-type star spectrum to clean from
        telluric absorptions.}
    \end{figure}

\subsection{Spectroscopy}

Spectroscopic observations of V505~Per were carried out along a nine-year
period (1998$-$2007) with the Echelle+CCD spectrograph mounted at the
Cassegrain focus of the 1.82~m telescope, operated by INAF Osservatorio
Astronomico di Padova on top of Mt. Ekar in Asiago (Italy). We obtained a
total of 36 spectra, well distributed in orbital phase and each one exposed
for 1800 sec. The latter corresponds to 0.5\% of the orbital period, so
introducing negligible smearing in orbital phase while allowing to get
excellent S/N. Its mean value over different Echelle orders is given in the
last column of Table~1 for all analyzed spectra. The median of these
S/N is 200.

Spectra up to January 2000 (HJD$\leq$2451567 in Table~1) were obtained with
the same instrument set-up as adopted for Paper~I: wavelength interval
4500$-$9480~\AA, resolving power $R$~$\sim$~20\,000,
UV~coated~Thompson~CCD with 1024$\times$1024 pixel (19 $\mu$m in size), and
a 2 arcsec slit East-West oriented. In 2004 (HJD$\geq$2454020) the detector
changed to a thinned EEV~CCD47-10 1024$\times$1024 pixel (13 $\mu$m in
size), covering the interval 3820$-$7290~\AA. The 2~arcsec slit East-West 
oriented and the $R$~$\sim$~20\,000 resolving power were maintained.

Data reduction was performed in a standard fashion with IRAF package running
under Linux operating system, including modeling and subtraction of the
scattered light. Great care was taken to ensure the highest quality in the
wavelength calibration for the highest accuracy in radial velocity
measurement.

Our Echelle spectrograph is attached at the Cassegrain focus, so it moves
with the telescope and it suffers from flexures. The latter are however
quite small in amplitude. These flexures are smooth and repeatable with
telescope pointing, and do not suffer from hysteresis as studied in detail
by Munari \& Lattanzi (1992). Each science exposure on V505 Per was
bracketed by exposures on the thorium comparison lamp, so to remove the
linear component of the flexure pattern. To check and compensate for the
presence of a (minimal) non-linear component of the flexure patter, the rich
telluric absorptions complexes at $\lambda\lambda$~5880-5940, 6275-6310,
6865-7050 and 7160-7330~\AA\ were cross-correlated on each V505~Per spectra
with a zero-velocity synthetic telluric absorption spectrum.  The typical
amount of radial velocity shift derived by this cross-correlation was
0.4~km~sec$^{-1}$. In no case the measured shifts exceeded
1.2~km~sec$^{-1}$, even for the spectra secured at large hour angles. These
shifts were than subtracted to the measured radial velocities so to remove
(at the 0.1~km~sec$^{-1}$ level) any instrumental pattern.

A check on the accuracy of the wavelength calibration of the scientific
spectra, after zero point correction for telluric lines, was carried out by
measuring the wavelength of night-sky lines, which are abundant in our
spectra. The largest contributors to the night sky lines are [OI] and OH,
with the addition of city lines due to HgI and NaI. Using as reference
wavelengths those given by Osterbrock et al. (2000), we verified a mean
velocity of night sky and city lines of 0.0~($\pm$0.1)~km~sec$^{-1}$ on each
science spectrum.

\section{Radial velocities}

Radial velocities were obtained following the same strategy as described in
Paper~I. We used the two-dimensional cross-correlation algorithm TODCOR  
described by Zucker \& Mazeh (1994). We applied it to 
the six adjacent Echelle orders (\#40-45) that cover the wavelength range
$\lambda\lambda~4890-5690$ \AA. These orders ($i$) lay close to the optical 
axis of the spectrograph where optical quality is the best, ($ii$) they are
densely packed by strong and sharp absorption lines, mainly due to FeI, 
and ($iii$) over them the S/N of the recorded spectra reaches peak values.
Before cross-correlation, each order was further trimmed so as to retain the
central 50\%, where the instrumental response and PSF sharpness is the best.

The appropriate template spectra for cross-correlation were selected among
the M05 synthetic spectra as those matching the $T_{\rm eff}$, $\log g$,
$V_{\rm rot}$~$\sin i$ and [M/H] of V505~Per components found by $\chi^2$
analysis (see Sect.~6 below).

Radial velocities obtained in the six trimmed orders were averaged and the
resulting mean value for each given spectrum, with the corresponding error
of the mean, are reported in Table~1.  The median value for the errors of 
the mean is $\sim$0.4~km~sec$^{-1}$ for both components.

As underlined by Latham et al. (1996), one of the key advantages of TODCOR
compared with conventional one-dimensional cross-correlation techniques is
that it greatly reduces the systematic errors in the radial velocities
caused by line blending. However, Torres et al. (1997, 2000) or Torres \&
Ribas (2002), found that systematic errors could survive when short
wavelength ranges (as in the case of the CfA speedometer they used) are
cross-correlated on fast rotating stars. We have verified that our radial
velocities are free from this systematic effect, thanks mainly to the wide
wavelength range and multi-Echelle-order type of our spectra and, to a
lesser extent, also to the slow rotation of the components of V505~Per. To
this aim, we have repeated the same test as carried out by Torres et al. 
(1997, 2000) or Torres \& Ribas (2002), namely we built synthetic binary
spectra by combining, with the proper luminosity ratio, the primary and
secondary templates, shifted to the appropriate velocities for each of the
actual exposure as predicted by the orbital solution. These synthetic binary
spectra were than fed to TODCOR in the same way as observed spectra were
fed previously and the resulting velocities were compared with the input
(synthetic) values. The differences (TODCOR minus synthetic), averaged over
the six Echelle orders, did not show a dependence with orbital phase and
never exceeded 0.2~km~sec$^{-1}$. A further test on the accuracy of TODCOR
radial velocities has been carried out by comparing them with the results of
radial velocities derived by measurement of individual lines fitted with
Gaussian profiles. We selected 42 of the strongest and more isolated
absorption lines and measured them for both components on all spectra
obtained close to quadratures (orbital phases 0.12$\leq$$\phi$$\leq$0.38 and
0.62$\leq$$\phi$$\leq$0.88). Again, the difference between TODCOR and
line-by-line radial velocities has no dependence on orbital phase and it is
always very small, with a r.m.s. value of 0.27~km~sec$^{-1}$.

    \begin{figure*}
    \centerline{\psfig{file=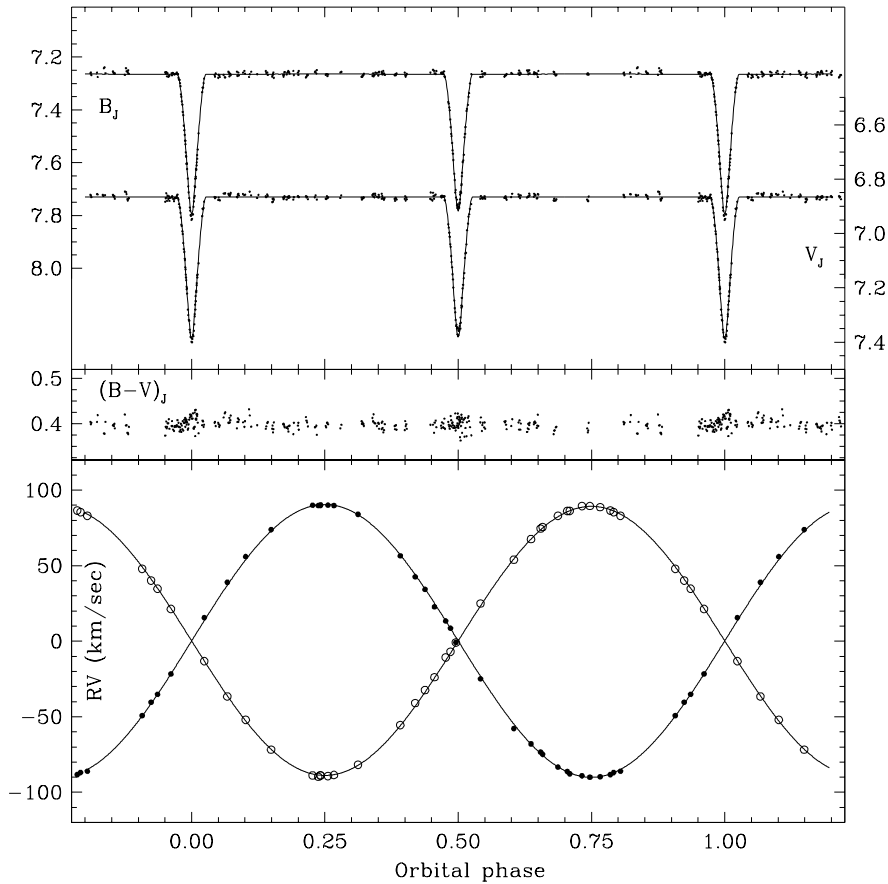,width=12.0cm}}
    \caption{Our $B_{\rm J}$, $V_{\rm J}$, $(B-V)_{\rm J}$ and
	radial velocity data of V505~Per. In the radial velocity panel,
	the open circles indicate the hotter and more massive (primary)
	star, while the filled circles pertain to the cooler and less
	massive (secondary) star. The orbital solution from Table~2 
        is over-plotted to the data.}
    \end{figure*}

\section{Interstellar reddening}

The amount of interstellar reddening is obviously critical to the
determination of absolute magnitude and therefore distance to the system. 
Our high-resolution spectra are ideally suited to search for and, if
detected, measure the intensity of interstellar absorption lines. 
In particular we searched for NaI (5890 \& 5896 \AA) and KI (7699.0 \AA) 
lines that Munari \& Zwitter (1997) showed to be excellent means to measure
the reddening. 

To cope with the presence of two overlapped spectra in our binary, we
searched for radial velocity constant NaI and KI lines at different orbital
phases (cf. Fig.~1).  For V505~Per we did not detect any such line with an
equivalent width exceeding 0.03~\AA, so that the reddening affecting
V505~Per is $E_{B-V}$$\leq$~0.01 mag. This corroborates the assumption
of a zero reddening by both M97 and M01.

\section{Analysis of stellar atmospheres}

Atmospheric parameters for the two components of V505~Per were obtained via
$\chi^2$ fitting to M05 synthetic spectral library. The $\chi^2$ fitting was
performed both on the single-lines spectrum obtained at phase 0.497 (cf.
Table~1), i.e. at the bottom of secondary eclipse, as well as on spectra at
quadratures, i.e. those showing the largest line split between the two
components.

The analysis of the 0.497 phase spectrum (vastly dominated by the light of
the primary star alone) provided $T_{\rm eff}$=6484$\pm21$~K, $\log
g$=4.25$\pm$0.07, [M/H]=$-$0.12$\pm0.03$ and $V_{\rm rot}\sin
i$=15.3$\pm$1.0 km~sec$^{-1}$.  These values are affected by a residual,
small fraction of light coming from the slightly cooler secondary star,
which passes behind the primary at secondary eclipse. We will later see from
orbital solution that the two components of the binary share very similar
$\log g$ and $V_{\rm rot}\sin i$. Therefore the only significant correction
to these $\chi^2$ values required by the residual contribution of the
secondary star to the combined system light at orbital phase 0.497 concerns
the effective temperature. We performed iterative orbital solutions and at
each iteration we derived the difference in the temperature between primary
and secondary star and the fraction of the combined system light due to the
two components. Convergence was reached for $T_{\rm eff}$=6512$\pm21$~K for
primary star, which was adopted in the orbital solution.  The atmospheric
parameters for the primary star derived from the phase 0.497 spectrum were
confirmed by $\chi^2$ analysis of the spectra obtained in quadrature, with
resulting parameters for the secondary star being $T_{\rm
eff}$=6460$\pm30$K, $\log g$=4.25$\pm$0.06, $V_{\rm rot} \sin
i$=15.4$\pm$1.0~km~sec$^{-1}$, and a metallicity [M/H]=$-$0.12$\pm0.03$
identical to that of the primary star. The small difference in temperature
between the two stars ($T_{\rm eff,1}$$-$$T_{\rm eff,2}$=52~K) is in
agreement with the similarly small difference in eclipses depths (0.04 mag)
displayed by both our and M97 photometry. The formal errors on $T_{\rm
eff,1}$ and $T_{\rm eff,2}$, 21 and 30~K respectively, are the error of the
mean computed on the results from independent $\chi^2$ analysis of each
individual Echelle spectral order. These errors are in line with typical
results of $\chi^2$ and minimum-distance-method analysis from high resolution, 
high S/N spectra (cf. Kovtyukh et al. 2006).

Comparison of V505~Per spectra with the high resolution spectral atlases of
Munari \& Tomasella (1999) and Marrese et al. (2003) indicates a similar
$\sim$F5V classification for the two components. The MK classification is a
discrete one, and the range of temperatures covered by the F5V box (i.e.
from F4.5V to F5.5V) is 6500$\leq$$T_{\rm eff}$$\leq$6650~K according to
recent calibration by Bertone et al. (2004). Similarly, the F5V box spans
the color range 0.425$\leq$$B$$-$$V$$\leq$0.455 following Popper (1980),
Straizys (1992) and Drilling and Landolt (2000). Our determination of 
V505~Per color is $B$$-$$V$=0.410$\pm$0.023. The 0.023 mag uncertainty 
is just the corresponding uncertainty in the Tycho $(B-V)_T$ color of the 
comparison star HD~14444, and does not include errors associated to the
Bessell (2000) transformation from Tycho to Johnson's system. It is worth
noticing that M97 reports a $B$$-$$V$=0.43 color for V505~Per (no
uncertainty associated to the transformation from local to Johnson's system
is provided). Thus, we may conclude that spectral classification and
photometric colors are mutually consistent and in agreement with results of
atmospheric analysis. The wide intervals in $T_{\rm eff}$ and color
associated to a spectral type, and the uncertainties in color transformation
from local to standard systems, argue in favor of our choice to adopt for
the orbital solution the temperature of the primary as derived by $\chi^2$
atmospheric analysis. The accuracy of the latter speaks by itself,
considering that ($a$) it provides the same surface gravity and $T_{\rm
eff,1}$$-$$T_{\rm eff,2}$ as derived independently by the orbital solution,
($b$) the same metallicity and derived by fitting to theoretical isochrones
and stellar evolutionary tracks, ($c$) the distance derived by adopting
$T_{\rm eff,1}$ from $\chi^2$ atmospheric analysis accurately match the
Hipparcos distance.

It is worth noticing that the comparison of V505~Per spectra with synthetic
ones and those of non-binary F5V field stars excludes the presence of
emission line cores or veiling in the CaII H \& K and far red CaII triplet
lines. This would be the case for chromospheric active and/or spotted
stellar atmospheres (e.g. Ragaini et al. 2003), of the type induced 
by differential rotation in synchronized close binaries (e.g. Munari 2003).
The wide orbital separation and slow synchronized velocities in V505~Per
are therefore in agreement with the lack of emission cores in CaII lines.
  
\section{Orbital solution}

\subsection{Orbital period stability}

Our photometry provides three epochs of minima: a primary eclipse at
HJD=2451587.30641 ($\pm$0.00017) and secondary eclipses at
HJD=2451779.40820 ($\pm$0.00016) and 2451910.29094 ($\pm$0.00017). We have
combined them with the orbital ephemeris in Table~2 and the epochs of minima
published by Kaiser et al (1990), M97 and
Demircan et al. (1997). The earliest of such minima dates back to 1903.
No appreciable O-C trend is found, which means that the derived period is
accurate and stable. 

\subsection{Initial parameters and modeling strategy}

A simultaneous spectroscopic and photometric solution for V505~Per was
obtained with the WD code (Wilson \& Devinney 1971, Wilson 1998) in its
WD98K93d version as developed by Milone et al. (1992), adopting MODE-2
option, appropriate for detached binary stars. As a starting point in the
iterative orbital solution, we imported the orbital parameters of M01 and
the atmospheric parameters derived by $\chi^2$ analysis.

Limb darkening coefficients were taken from van Hamme (1993) for the
appropriate metallicity, temperature and gravity. A linear law for limb
darkening is usually assumed for convective atmospheres (as it is the case
for F5V stars). Nevertheless, for check and completeness, we re-run full
orbital solutions also with logarithmic and square--root limb darkening laws
as well as for metallicities [M/H]=$-1.0$, $-0.5$, +0.0 and +0.5. The
response of orbital solution to these changes was minimal, with orbital
parameters not varying by more than their (quite small) formal errors. For
final solution we retained the [M/H]=$-$0.12 metallicity derived by the
$\chi^2$ atmospheric analysis and a linear limb darkening law. The final
adopted solution (cf. Table~2) converged to the following limb darkening 
parameters (in the WD jargon): $x_{bol,1}$=0.660, $x_{bol,2}$=0.666, 
$x_{V,1}$=0.425, $x_{V,2}$=0.426, $x_{B,1}$=0.521, $x_{B,2}$=0.522.

Driven by the lack of evidence for multiple reflection effects in the
V505~Per lightcurve presented in Fig.~2, we run the final orbital solution
including only the inverse square law illumination. For sake of completeness,
orbital solution tests were carried out including also multiple reflections.
Their inclusion did not improve the accuracy and convergence solution and
therefore were considered no further.

Bolometric albedos ($A_{1,2}$) and exponents in the bolometric gravity
brightening law ($g_{1,2}$) were set to 0.5 and 0.3, respectively, as are
expected for convective envelopes (cf. Wilson 1998). Orbital solutions were
carried out testing various combinations of these parameters in the range
$0.5\leq$$A_{1,2}$$\leq 1.0$ and $0.3\leq$$g_{1,2}$$\leq 1.0$ ($A$ and $g$
are expected to be unity for radiative envelopes), without improvement in
the accuracy.
  
\subsection{The orbit}

   \begin{table}
   \caption{Orbital solution for V505~Per (over-plotted to observed data in
    Fig.~2). Formal errors to the solution are given. The last two lines
    compare the {\it original} and {\it re-computed} Hipparcos trigonometric
    parallax and its 1$\sigma$ error (see text) with the distance
    we derived from the orbital solution in the $E_{B-V}$=0.00 case and
    $E_{B-V}$=0.01 upper limit to the reddening affecting V505~Per.}
    \begin{center}
    \centerline{\psfig{file=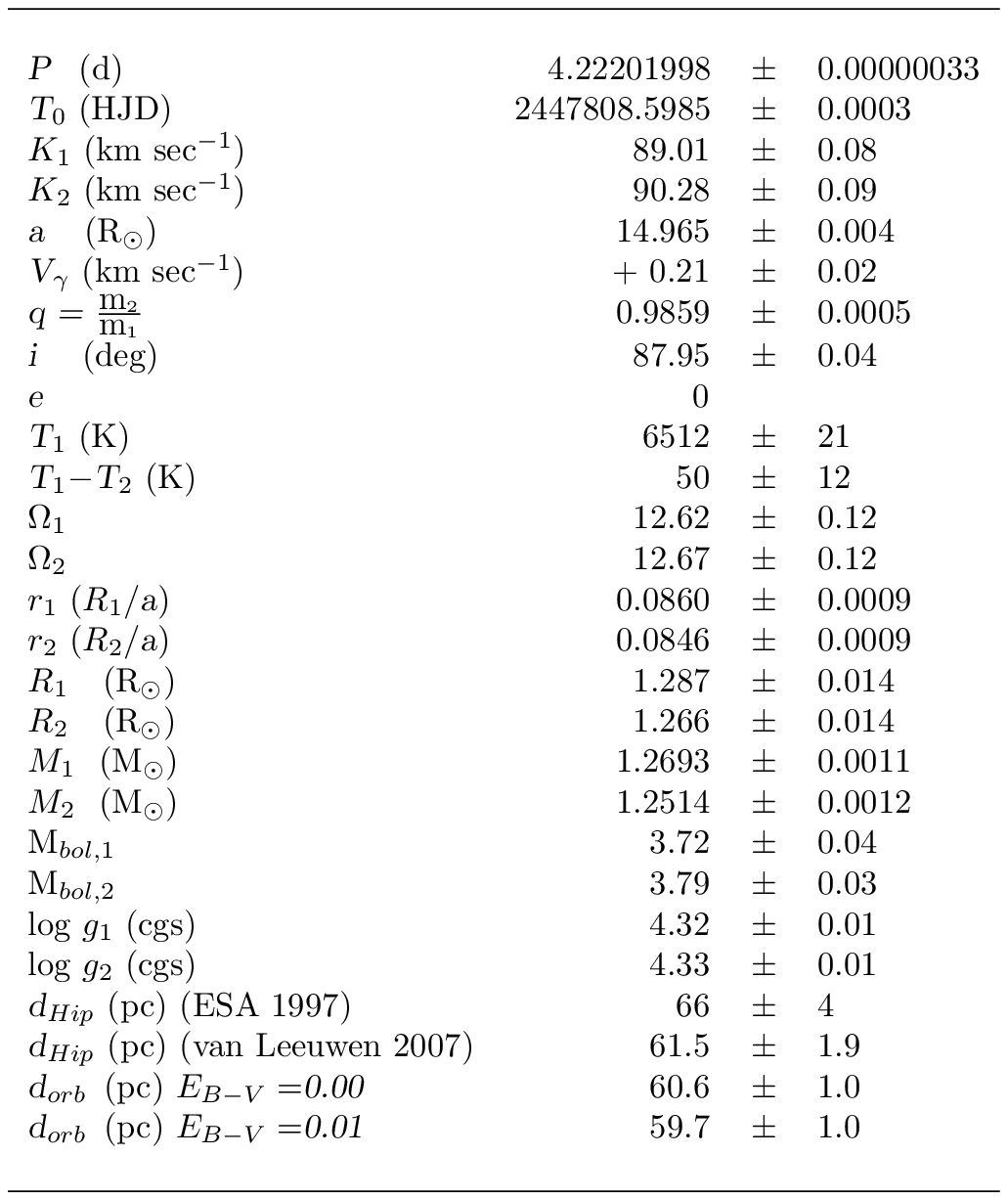,width=8.5cm}}
    \end{center}
    \end{table}

The final orbital solution we converged upon for V505~Per is presented in
Table~2 and over-plotted to our observational data in Fig.~2. The r.m.s.  of
photometric data with respect to the orbital solution is 0.008 and 0.007 mag
for $B_{\rm J}$ and $V_{\rm J}$ photometry, respectively. The r.m.s. of the
measured radial velocities with respect to the orbital solution is 0.68 and
0.84~km~sec$^{-1}$ for primary and secondary star, respectively.

The final orbital solution  rests on our radial velocities and photometry,
augmented by inclusion of M97 photometry during eclipses (i.e. M97
photometric points from 0.95 to 0.05 and from 0.45 to 0.55 orbital phases)
that helps to fill in the gaps in our lightcurves at ingress and egress
from both eclipses.  Inclusion of this subset of M97 data significantly
reduced the formal errors on both stellar radii without altering the orbital
solution. Furthermore, to the benefit of better constraining $P$ and
$t_{0}$, we imported the epoch of minimum given in M97 (HJD=2447808.5998
$\pm$ 0.0001).  We did not include M97 radial velocities and 
photometry outside eclipse phases because they would not improve the
convergence of the orbital solution and instead widen the formal errors (due
to lower accuracy of M97 photometry and radial velocities, which are
characterized by a dispersion of 0.01 mag in both photometric bands and
1.0~km~sec$^{-1}$ for both radial velocity curves with respect to the
orbital solution). We also attempted orbital solutions by considering only
$B$ or $V$ photometric data. This led to very similar orbital solutions,
differing much less than their formal errors. They both also well include in
their error bars the adopted final orbital solution.

\subsection{Physical parameters}

All system parameters are well constrained by the orbital solution in
Table~2, particularly those most dependent on the radial velocities (i.e.
$a$, $M_{\rm 1}$, $M_{\rm 2}$, $q$ and $V_{\gamma}$). Formal accuracies are
0.09 and 0.10\% on the masses, and 1.1 and 1.0\% on the radii. We found no
evidence for a non-spherical shape, as $R_{pole}$/$R_{point}$=1.00 for both
stars.

The synchronized rotational velocities of the two stars would be
15.4~km~sec$^{-1}$ and 15.2~km~sec$^{-1}$. They are well within the error
bars of the results of the $\chi^2$ fit: 15.3$\pm$0.5~km~sec$^{-1}$ and=
15.4$\pm$1.0~km~sec$^{-1}$ (see Sect.~6). We therefore conclude that both
components of the binary rotate synchronously with the orbital motion.

 \begin{figure}
 \centerline{\psfig{file=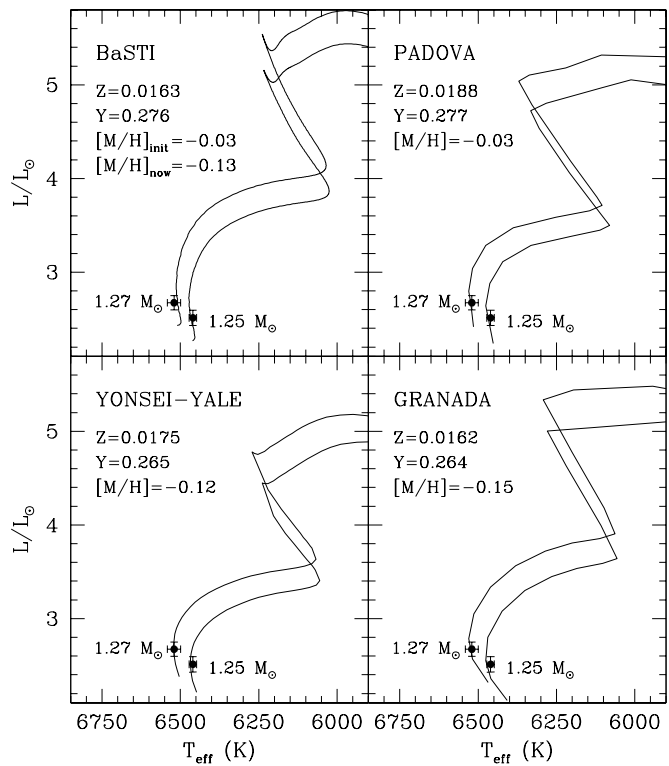,angle=0,width=8.8cm}}
 \caption{Comparison of evolutionary tracks from various libraries (see the
 text for more details), with the observed parameters for the
 components of the binary V505 Per (filled circles). For the best fitting
 evolutionary tracks, the values of the metallicity and helium content 
 are given. The BaSTI tracks have been computed on purpose for this paper
 for exactly the masses of the two components (1.2693 and 1.2514
 $M_\odot$).}
 \end{figure}

Surface gravities and $T_{\rm eff,1}$$-$$T_{\rm eff,2}$ from the orbital
solution ($\log g_{1}$=4.32$\pm$0.01, $\log g_{2}$=4.33$\pm$0.01,
$T_{\rm eff,1}$$-$$T_{\rm eff,2}$=50~K) are in excellent agreement with
those from $\chi^2$ atmospheric fit ($\log g_{1}$=4.25$\pm$0.07, $\log
g_{2}$=4.25$\pm$0.06 $T_{\rm eff,1}$$-$$T_{\rm eff,2}$=52~K).

\begin{figure}
 \centerline{\psfig{file=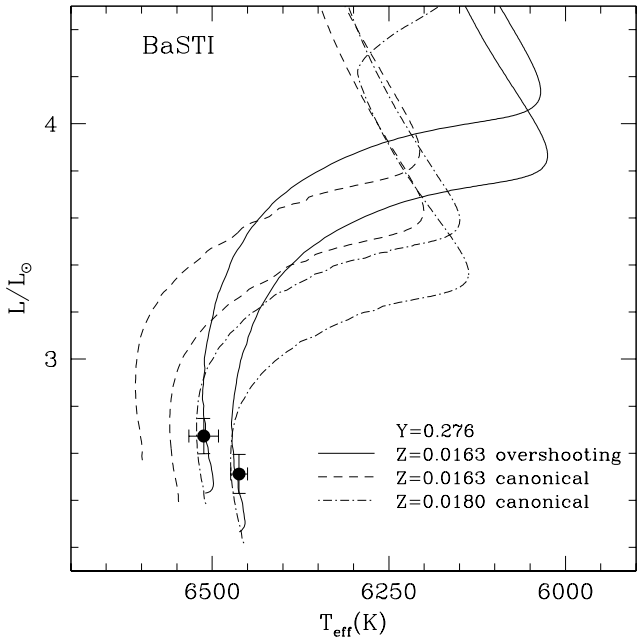,width=8.8cm}}
  \caption{As in Fig.~3, but only for the BaSTI models. Selected evolutionary
  models computed by neglecting the occurrence of convective core
  overshooting are also shown. The adopted initial metallicities are
  labeled.}
\end{figure}

\subsection{Distance to the system}

To compute a distance to V505~Per from the orbital solution we adopted, from
Bessell, Castelli and Plez (1998), a bolometric correction $BC$=0.00 for
both components and a bolometric magnitude for the Sun $M_{\rm
bol,\odot}$=4.74. We also assumed a standard $A_{V}$=$E_{B-V}$$\times$~3.1
reddening law and a color excess $E_{B-V}$$\leq$~0.01, as discussed in
Sect.~5. The corresponding distance is 60.6$\pm$1.0 for $E_{B-V}$=0.00, and
59.7$\pm$1.0 for $E_{B-V}$=0.01 (cf. Table~2). The Hipparcos (ESA 1997)
distance to V505~Per is given as 66$\pm$4 pc. Recently, van Leeuwen \&
Fantino (2005) and van Leeuwen (2007) have performed a re-reduction of the
satellite data and produced a revised Hipparcos astrometric catalog.  The
revised Hipparcos distance to V505~Per (kindly communicated to us by F.
van Leeuwen in advance of publication) is 61.5$\pm$1.9 pc, in better
agreement with our orbital solution than original Hipparcos distance.

The literature lacks consensus for the bolometric magnitude of the Sun, the
listed values spanning from $M_{\rm bol,\odot}$=4.72 (Lang 2006) to $M_{\rm
bol,\odot}$=4.77 (Bowers \& Deeming 1984, see also Allen 1976,
B\"ohm-Vitense 1984 and Zombeck 1990). The effect of adopting them would be
to shift by $\pm$1~pc the distance derived from the orbital solution. The
scatter in literature in even larger for the bolometric correction. 
Actually, although its definition is a straightforward one, there is some
confusion resulting from the choice of zero-point, as pointed out by
Bessell, Castelli and Plez (1998). For guidance, the V505~Per distance would
decrease by less than 1~pc assuming $BC$=$-0.03$ from Popper (1980).

\section{Comparison with theoretical stellar models}

One of the most important issue of current stellar models is the actual
extension of the convective core during the central H-burning stage: how
much larger the convective core extension is with respect to the canonical
prediction provided by the classical Schwarzschild criterion, i.e. the
amount of convective core overshooting (cf. Cassisi 2004 and references
therein). 

The study of open clusters has shown that stellar models must
allow for the occurrence of convective core overshooting in order to provide
a satisfactory match to the observed CMDs (Kalirai et al. 2001 and
references therein).  On the other hand, a still unsettled issue concerns how
much the convective core overshooting reduces with decreasing of stellar
mass (cf. Woo \& Demarque 2001 for a detailed discussion). 

Among open clusters, only those with a {\em turn-off} (TO) mass
$\sim$1.2~M$_\odot$ are useful tests in this respect, with M67 being the
main target of current investigations, in particular those of Sandquist
(2004) and Vandenberg \& Stetson (2004). Their main conclusion - also
supported by the results shown by P04 - was that the comparison between the
CMD of M67 and theoretical isochrones seems to indicate that the extension of
the overshooting region has to be already almost down to zero for masses
$\sim$1.2~M$_\odot$, at least for metallicities around solar. Due to current
uncertainties in empirical estimates about clusters distance, reddening and
heavy elements abundances (cf. Vandenberg et al. 2007), this result on M67
needs independent confirmations.

In this context a relevant contribution can be provided by
binary systems whose components have a suitable mass. This approach has been
been already adopted in literature for some well-studied binaries, namely
V459~Cas (Sandberg Lacy et al. 2004), TZ~For (Vandenberg et al. 2006, and
references therein) and AI~Phe (Andersen et al. 1988; P04, and references 
therein). The masses above derived for the V505~Per components qualify, in
principle, this binary as a suitable system for testing the efficiency of
convective core overshooting.

\begin{figure}
 \centerline{\psfig{file=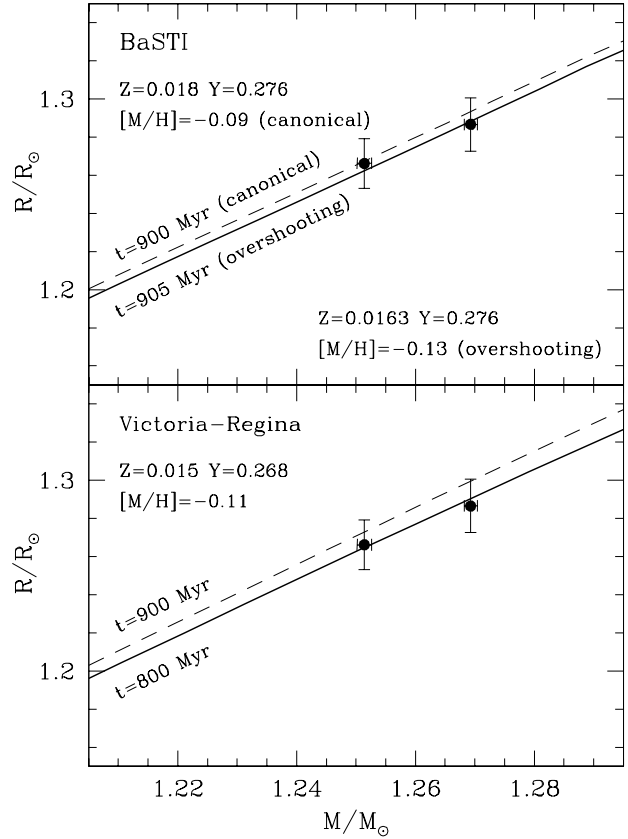,width=8.5cm}}
  \caption{{\sl Top panel}: Comparison between the observed masses and
  radii of the components of the eclipsing binary V505 Per (filled circles and
  $1\sigma$ errors), and the predictions of both an overshooting and a
  canonical isochrone from the BaSTI library, for the labelled choices about
  the metallicity and age. {\sl Bottom panel}: as above but for
  overshooting isochrones provided by the Victoria-Regina library.}
\end{figure}

In Fig.~3 we compare on the H-R diagram the position of the two
components of V505~Per with some of the most widely used and accurate
libraries of stellar tracks and isochrones: the BaSTI models as provided by
P04 (see also Cordier et al. 2007), the Padova models as published by
Girardi et al. (2000), the Yonsei-Yale isochrones provided by Yi et al.
(2001) in their second release, and the Claret et al. (2003) Granada
library. All such libraries of evolutionary models have been computed by
assuming a scaled-solar heavy elements distribution (we refer to the quoted
papers for details about the adopted solar heavy elements mixtures).

The tracks plotted in Fig.~3 (with the exception of BaSTI, see below) have
been obtained by linear interpolation of the published tracks closest in
mass to those of the two V505~Per components. For each library of stellar
models, the reported metallicity is the one best fitting the position of the
two V505~Per components.  Concerning BaSTI tracks in Fig.~3, we computed
on purpose for this paper a grid of models for exactly the masses of the two
components (1.2693 and 1.2514 M$_\odot$), varying the helium and heavy
elements content. The tracks were computed both in the standard canonical
scenario with no convective core overshooting as well as including it, in
which case we adopted the same prescriptions discussed by P04. The best
fitting metallicity for BaSTI models is [M/H]=$-$0.13, almost exactly the
same as derived by the spectroscopic atmospheric analysis ([M/H]=$-$0.12,
cf. Sect. 6). All BaSTI models computed for this paper properly account (see
P04) for both helium and heavy elements diffusion.

Element diffusion in stellar interiors is a relevant issue that can confuse
comparison between observed values and theoretical predictions. During the
lifetime of a star, heavier nuclei tend to sink deeper within the star while
lighter ones migrate toward the surface. Thus, a star formed homogeneous,
progressively decreases its surface metallicity while living on the main
sequence. The BaSTI models, computed including element diffusion, show that
the initial [M/H]=$-$0.03 metallicity decreases to [M/H]=$-$0.13 at stellar
surface for the $\sim$0.9~Gyr age of V505~Per. The Granada and Padova tracks do
not account for the element diffusion. Their [M/H] values in Fig.~3
pertain therefore to model of initial homogeneous composition, kept constant
during model computations. On the other hand, the Yonsei-Yale tracks account
for He diffusion but not for metals diffusion. In spite of the unknown
relation between initial and current metallicity for Padova, Granada and
Yonsei-Yale tracks, their ability to reproduce the basic physical parameters
measured for V505~Per is nevertheless quite reasonable. 

Fig.~3 shows also that V505 Per system is a not-significantly evolved
binary systems: the secondary star is still near to its Zero Age Main
Sequence location, while the primary component is only slightly more
evolved.

It is worth noticing that all the evolutionary tracks plotted in Fig.~3
account for a very small amount of convective core overshooting (see quoted
references for more details about the overshooting efficiency adopted by the
various authors). To investigate the role of overshooting in greater detail,
we computed a grid of BaSTI models for exactly the masses of the two
components, with element diffusion but with null overshooting (cf. Fig.~4).
These models require a larger initial metallicity ($Z=0.0183$) to fit the
position of V505~Per components, and they predict a current surface
metallicity [M/H]=$-$0.09 for V505~Per. Even if marginally within the error
bar, this value is offset with respect to observed metallicity. More
important, these canonical models without overshooting are not able to
provide a similar age to both components of V505 Per: the age difference
between the stellar models fitting the two stars is $\sim$0.11 Gyr. On the
other hand, BaSTI stellar models that include overshooting fit very
well the location of both V505~Per components in the H-R diagram, their
current metallicity is closer to observed value, and their age difference at
the fitting point is of about 0.05 Gyr. Therefore, we conclude that, at the
masses of the V505~Per components, some convective core overshooting must be
present.

To quantify its exact amount, we computed BaSTI models for different
efficiencies of the convective core overshooting. The efficiency is usually
defined in terms of the parameter $\lambda_{\rm OV}$ that gives the length -
expressed as a fraction of the local pressure scale height $H_{\rm P}$ -
crossed by the convective cells in the convective stable region outside the
Schwarschild convective boundary. The BaSTI models providing the best fit to
observations have efficiencies $\lambda_{\rm OV}=0.093$ and 0.087, for the
1.2693~M$_\odot$ primary component and its 1.2514~M$_\odot$ companion,
respectively. This suggests that stars with masses $\sim$1.2~M$_\odot$ have
an overshooting region already very small, supporting the conclusions
obtained by P04 for the system AI Phe and by Vandenberg et al. (2006 and
references therein) for the stars belonging to the cluster M67. It is worth
to note that even if coarsely similar, the mass of the two components of AI
Phe are lighter than those of V505 Per: 1.231+1.190 vs 1.269+1.251
M$_\odot$. Both pairs are in the critical range where the efficiency of
convective overshooting changes with the stellar mass\footnote{ Regardless
of the initial metallicity, P04 adopt $\lambda_{\rm OV}$=0.20$\times$$H_{\rm
P}$ for $M$$\geq$1.7~M$_\odot$, $\lambda_{\rm OV}$=0 for 
$M$$\leq$1.1~M$_\odot$, and $\lambda_{\rm OV}$=($\frac{M}{M_\odot}$ $-$ 0.9)/4 for 1.1
M$_\odot\leq$M$\leq$ 1.7 M$_\odot$}.  So it could really be the case that
overshooting is already null for the AI Phe masses and very small, but not
negligible, for V505 Per ones.

Before concluding, it is worth noticing that SB2 EBs with a larger
difference in mass between the two components and older than V505~Per would
be an even more stringent test of the amount of convective core
overshooting. In fact, both components of V505 Per are still close to their
zero-age main sequence, when the stellar radius does not depend largely on
the stellar age, which limits the use of the Radius - Mass diagram for
constraining the evolutionary scenario. This is shown in Fig.~5. The
comparison between the position of the two components of V505 Per with BaSTI
isochrones computed with diffusion and overshooting provides a best match
for an age of 0.905 Gyr, but a similar good fit can be obtained by adopting
canonical models for a slightly lower age, although at the fitting point in
the H-R diagram the two stellar models show a significant difference in age.
In Fig.~5 we have tried to similarly estimate the age of V505~Per using also
the Victoria-Regina library of stellar models provided by Vandenberg et al.
(2006) that accounts for convective core overshooting but neglect the
occurrence of diffusive processes. In this case the age of V505~Per would be
$\sim0.80$~Gyr.

\begin{acknowledgements}
We would like to thank the skillful assistance by the technical staff
operating the 1.82m telescope in Asiago, L.A. Marschall and D.B. Williams
for kindly communicating in electronic form their published observational
data, F. van Leeuwen for its revised Hipparcos parallax of V505~Per prior to
publication, and M. Valentini and R. Barbon for useful discussions. We
warmly thanks D. Vandenberg for providing us his own evolutionary models.
\end{acknowledgements}


\begin{thebibliography}{}
\bibitem[1976]{allen}     Allen, C.W. 1976, in {\em Astrophysical Quantities} The Athlone Press
\bibitem[1988]{anders88}  Andersen, J., Clausen, J.V, Gustafsson, B., Nordstr\"om, B., Vanderberg, D.A. 1988, A\&A 196, 128
\bibitem[1991]{anders91}  Andersen, J. 1991, A\&AR, 3, 91
\bibitem[1997]{anders97}  Andersen, J. 1997, in {\em Fundamental Stellar Properties: The Interaction Between Observation
                          and Theory}, T. Bedding, A. Booth and J. Davis eds., Kluwer
                          Dordrecht, IAU Sym. No. 189, 99 
\bibitem[1998]{anders02}  Andersen, J. 2002, in {\em Observed HR Diagrams and Stellar Evolution} 
                          T. Lejeune and J.  Fernande eds., ASP Conf. Proc. 274, 187
\bibitem[1998]{bessell}    Bessell, M.S., Castelli, F., Plez, B. 1998, A\&A, 333, 231
\bibitem[2000]{bessell00} Bessell, M.S. 2000, PASP, 112, 961
\bibitem[2004]{bertone}  Bertone, E., Buzzoni, A., Chavez, M., Rodriguez-Merino, L.H. 2004, AJ, 128, 829
\bibitem[1989]{bohm}      B\"ohm-Vitense, E. 1989, in {\em Stellar Astrophysics}, Cambridge University Press
\bibitem[1984]{bowers}    Bowers, R.L., Deeming, T. 1984, in {\em Astrophysics}, Jones \& Bartlett Publishers
\bibitem[2004]{cassisi04} Cassisi, S. 2004, in {\em Variable Stars in the Local Group}, IAU Colloquium 193, 
Christchurch (New Zealand), D. W. Kurtz and K. R. Pollard eds., ASP Conf. Proceedings, Vol. 310, 489
\bibitem[2003]{claret3}   Claret, A., Paunzen, E., Maitzen, H.M. 2003, A\&A, 412, 91 
\bibitem[2007]{cordier}    Cordier, D., Pietrinferni, A., Cassisi, S., Salaris, M. 2007, AJ, 133, 468
\bibitem[1997]{demircan}  Demircan, O., Ozdemir, S., Tanriver, M. 1997, Ap\&SS 250, 327
\bibitem[2000]{landolt}   Drilling, J.S., Landolt, A.U. 2000, in Allen's Astrophysical Quantities, A.N.Cox ed., Springer, 381
\bibitem[1997]{hipp}      ESA 1997, {\em The Hipparcos and Tycho Catalogues}, ESA SP-1200
\bibitem[2000]{girardi}   Girardi, L., Bressan, A., Bertelli, G., Chiosi, C. 2000,  A\&AS 141, 371
\bibitem[1989]{kaiser}     Kaiser, D.H. 1989, AAVSO, 18, 149
\bibitem[1990]{kaiser2}    Kaiser, D.H., Balwin, M.E., Williams, D.B. 1990 IBVS No. 3442
\bibitem[2001]{kalirai} Kalirai, J. S., Richer, H. B., 
Fahlman, G.G., Cuillandre, J.C., Ventura, P., D'Antona, F., Bertin, E., Marconi, G., Durrell, P.R. 2001, AJ, 122, 266
\bibitem[2006]{kovtyukh}  Kovtyukh, V.V., Soubiran, C., Bienayme, O., Mishenina, T.V., Belik, S.I. 2006, MNRAS 371, 879
\bibitem[2006]{lang}      Lang, K.R. 2006, in {\em Astrophysical Formulae}, 3rd edition, Springer
\bibitem[1996]{la96}  Latham, D. W., Nordstr\"om, B., Andersen, J.,
Torres, G., Stefanik, R. P., Thaller, M., Bester, M. J. 1996, A\&A, 314, 864
\bibitem[2003]{marrese}   Marrese, P.M., Boschi, F., Munari, U. 2003, A\&A, 406, 995
\bibitem[1990]{m90}       Marschall, L.A., Stefanik, R.P., Nations, H.L., Davis, R.J. 1990 IBVS No. 3447
\bibitem[1997]{m97}       Marschall, L.A., Stefanik, R.P., Lacy, C.H., Torres, G., Williams, D.B., Agerer, F. 1997, AJ, 114, 793 (M97)
\bibitem[1992]{milone}    Milone, E.F., Stagg, C.R., Kurucz, R.L. 1992, ApJS 79, 123
\bibitem[1992]{munari92}  Munari, U., Lattanzi, M.G. 1992, PASP, 104, 121 
\bibitem[2003]{m03}   Munari, U. 2003, in ASP Conf. Ser. 298, 227
\bibitem[1997]{munari97}  Munari, U., Zwitter, T. 1997, A\&A, 318, 269
\bibitem[1999]{munari99}  Munari, U., Tomasella, L. 1999, A\&AS, 137, 521
\bibitem[2001]{munari01}  Munari, U., Tomov, T., Zwitter, T., Milone, E.F., Kallrath, J., Marrese, 
P.M., Boschi, F., Prsa, A., Tomasella, L., Moro, D. 2001, A\&A, 378, 477 (M01)
\bibitem[2005]{munari05}  Munari, U., Sordo, R., Castelli, F., Zwitter, T. 2005, A\&A, 442, 1127
\bibitem[2000]{osterb}    Osterbrock, D.E., Waters, R.T., Barlow, T.A., Slanger, T.G., Cosby, P.C. 2000, PASP, 112, 733
\bibitem[]{cassisi}        Pietrinferni, A., Cassisi, S., Salaris, M., Castelli, F. 2004, ApJ, 612, 168 (P04)
\bibitem[1997]{pols}       Pols, O.R., Tout, C.A., Schr\"oder, K.P., Eggleton, P.P., Manners, J. 1997, MNRAS, 289, 869  
\bibitem[1980]{popper}     Popper, D.M. 1980, ARA\&A 18, 115
\bibitem[2003]{ra03}       Ragaini, S., Andretta, V., Gomez, M. T., Terranegra, L., Bus\'a, I., Pagano, I. 2003, in ASP Conf. Ser. 298, 461
\bibitem[2000]{ribas00}    Ribas, I., Jordi, C., Gim\'enez, A. 2000, MNRAS, 318, L55
\bibitem[]{} Sandberg Lacy, C.H., Claret, A., Sabby, J.A. 2004, AJ, 128, 132
\bibitem[2004]{sandquist} Sandquist, E. L. 2004, MNRAS, 347, 101
\bibitem[2004]{siviero}   Siviero, A., Munari, U., Sordo, R., Dallaporta, S., Marrese, P.M., Zwitter, T., Milone, E.F.
                          2004, A\&A, 417, 1083 (Paper I)
\bibitem[2006]{steinmetz} Steinmetz, M., Zwitter, T., Siebert, A., Watson, F. G. and 50 coauthors 2006, AJ, 132, 1645  
\bibitem[1981]{straizys81}  Strai\v zys, V., \& Kuriliene, G. 1981, Ap\&SS 80, 353
\bibitem[1992]{straizys92}  Strai\v zys, V. 1992, Multicolor Stellar Photometry, Pachart Publishing House, Tucson
\bibitem[1997]{to97}        Torres, G., Stefanik, R.P., 
Andersen, J., Nordstr\"om, B., Latham, D.W., Clausen, J.V. 1997, AJ, 114, 2764
\bibitem[2000]{to00} Torres, G., Andersen, J., Nordstr\"om, B., Latham, D.W.
2000, AJ 119, 1942
\bibitem[2002]{to02} Torres, G., Ribas, I. 2002, ApJ, 567, 1140
\bibitem[2006]{Vandenberg06} Vandenberg, D.A., Bergbush, P. A., Dowler, P.D. 2006, ApJS, 162, 375
\bibitem[2007]{Vandenberg07} Vandenberg, D.A., Gustafsson, B., Edvardsson, B., Eriksson, K., Ferguson, J. 2007, ApJ Letter, {\sl in press}
\bibitem[2004]{Vandenberg04} Vandenberg, D.A., Stetson, P.B. 2004, PASP, 116, 997
\bibitem[1993]{vanhamme}  van Hamme, W. 1993, AJ, 106, 2096
\bibitem[2005]{vanl05}    van Leeuwen, F., Fantino, E. 2005, A\&A, 439, 791
\bibitem[2007]{vanl07}    van Leeuwen, F. 2007, {\em Hipparcos, the new reduction}, Springer, in press
\bibitem[1971]{wilson}    Wilson, R.E., Devinney, E.J. 1971, ApJ, 166, 605 
\bibitem[1998]{wilson2}   Wilson, R.E. 1998, {\it Computing Binary Star Observables}, Univ. of Florida Astronomy Dept.
\bibitem[2001]{wd} Woo, J.-H., Demarque, P. 2001, AJ, 122, 1602
\bibitem[2001]{Yi} Yi, S., Demarque, P., Lejeune, T., Barnes, S. 2001, ApJS 136, 417
\bibitem[1990]{zombeck}   Zombeck, M.V. 1990, in {\em Handbook of Space Astronomy \& Astrophysics}, 2nd edition, Cambridge University Press
\bibitem[1994]{zucker}    Zucker, S., Mazeh, T. 1994, ApJ, 420, 806
\bibitem[2004]{zwitter2}  Zwitter, T., Castelli, F., Munari, U. 2004, A\&A, 417, 1055 
\end{thebibliography}
\end{document}